\documentclass[aps,prc,twocolumn,showpacs,superscriptaddress,floatfix,10pt]{revtex4}
\usepackage{bm}
\usepackage{amssymb}
\usepackage{amsmath}
\usepackage{graphicx}
\usepackage{color}
\begin{document}

\title{Charged current neutrino interactions in core-collapse supernovae in a virial expansion}

\author{C. J. Horowitz}
\email{horowit@indiana.edu}
\affiliation{
Department of Physics and CEEM, Indiana University, Bloomington, Indiana 47405, USA
}%
\author{G. Shen}
\email{gshen@u.washington.edu}
\affiliation{Institute for Nuclear Theory, University of Washington, Seattle, WA 98195, USA}

\author{Evan O'Connor}
\affiliation{TAPIR, Mailcode 350-17, California Institute of Technology,
Pasadena, CA 91125, USA}
\affiliation{Canadian Institute for Theoretical Astrophysics, 60 St.~George Street, University of Toronto, ON M5S 3H8, Canada}
\author{Christian D. Ott}
\affiliation{TAPIR, Mailcode 350-17, California Institute of Technology,
Pasadena, CA 91125, USA}

\date{\today}

\begin{abstract}
Core-collapse supernovae may depend sensitively on charged current neutrino interactions in warm, low density neutron rich matter.  A proton in neutron rich matter is more tightly bound than is a neutron.  This energy shift $\Delta U$ increases the electron energy in $\nu_e+n\rightarrow p + e$, increasing the available phase space and absorption cross section.  Likewise $\Delta U$ decreases the positron energy in $\bar\nu_e+p\rightarrow n + e^+$, decreasing the phase space and cross section.  We have calculated $\Delta U$ using a model independent virial expansion and we find $\Delta U$ is much larger, at low densities, than the predictions of many mean field models.  Therefore $\Delta U$ could have a significant impact on charged current neutrino interactions in supernovae.  Preliminary simulations of the accretion phase of core-collapse supernovae find that $\Delta U$ increases $\bar\nu_e$ energies and decreases the $\nu_e$ luminosity.

\end{abstract}

\bigskip

\maketitle

\section{Introduction}
\label{sec.introduction}

Neutrino and antineutrino capture reactions play a crucial role in core collapse supernovae. They provide important heating that may reenergize a shock and lead to the explosion.   In addition, these reactions help determine the spectra of radiated $\nu_e$ and $\bar\nu_e$ that is important for detection of supernova neutrinos on earth \cite{Scholberg:2012id} and for neutrino oscillations \cite{Duan:2010,Raffelt:2010}.  Finally the rates of neutrino and antineutrino captures determine the ratio of neutrons to protons in the neutrino driven wind above a proto-neutron star.  This is crucial for nucleosynthesis, see for example \cite{Arcones:2012,Qian:2003,Horowitz:2002}.  

Supernova neutrinos are radiated from the neutrino-sphere, a warm low density gas of neutron rich matter, with a temperature near $T\approx 5$ MeV and a density of order $10^{-4}$ to $10^{-3}$  fm$^{-3}$.  Recently Roberts \cite{Roberts:2012zza} and Roberts and Reddy \cite{Roberts:2012um} have suggested that strong interaction energy shifts in this gas increase the cross section for 
\begin{equation}
\nu_e+n\rightarrow p+e\, ,   
\label{Eq.nue}
\end{equation}
by increasing the available phase space, and decrease the cross section for 
\begin{equation}
\bar\nu_e + p \rightarrow n + e^+\, .
\label{Eq.nubar}
\end{equation}
These shifts arise because a proton is more tightly bound in neutron rich matter than is a neutron and this binding energy difference increases the energy of the $e$ in Eq.~\eqref{Eq.nue} and decrease the energy of the $e^+$ in Eq.~\eqref{Eq.nubar}.   Martinez-Pinedo et al. \cite{MartinezPinedo:2012rb} have also performed supernova simulations with this energy shift.  However, these works \cite{Roberts:2012zza,Roberts:2012um,MartinezPinedo:2012rb} all calculate the shift in a mean field approximation that is very model dependent at low densities.  Therefore the size of the effect may be poorly determined.  

The energy shift 
\begin{equation}
\Delta U=U_n-U_p
\label{Eq.dU}
\end{equation}
is the difference in potential energy of a neutron $U_n$ or a proton $U_p$ in the medium.  This is closely related to the symmetry energy that describes how the energy of nuclear matter rises as one moves away from equal numbers of neutrons and protons.  The symmetry energy has been calculated at low densities using a virial expansion and found to be large because of correlations and the formation of bound states such as $^4$He, $^3$He, and $^3$H \cite{Horowitz:2005nd}.  Note that these light nuclei may also impact neutrino spectra \cite{O'Connor:2007eb,Arcones:2008kv}.

Warm low density nuclear matter can be produced in the laboratory with heavy ion collisions.  Kowalski et al. studied collisions of 35 MeV/nucleon $^{64}$Zn projectiles with $^{92}$Mo and $^{197}$Au target nuclei and inferred that the symmetry energy at densities of 0.01 to 0.05 times saturation density ranged from 9 to 13.6 MeV \cite{Kowalski:2007}, see also \cite{Natowitz:2010}.  These large values are in good agreement with virial expansion results.  However, they are much larger than the predictions of many mean field models, see for example refs. \cite{Roberts:2012um,IUFSU}.

In this paper we calculate the energy shifts and in medium cross sections for the reactions in Eqs.~\eqref{Eq.nue},\eqref{Eq.nubar} using a virial expansion.  The virial expansion makes model independent predictions for thermodynamic quantities based on elastic scattering phase shifts.  Some earlier virial expansion results include the equation of state of pure neutron matter \cite{Horowitz:2005zv}, the equation of state of nuclear matter including protons, neutrons, and alpha particles \cite{Horowitz:2005nd}, and the long wavelength neutrino response of pure neutron matter \cite{Horowitz:2006pj}.   Recently, full astrophysical equations of state giving the pressure of nuclear matter as a function of temperature, density, and proton fraction have been developed that reduce to the virial expansion at low densities \cite{Shen:2011b, Shen:2011a, Shen:2010}.

This paper is organized as follows.  We describe the virial expansion formalism in Sec. \ref{sec.formalism}.  Next, in Sec. \ref{sec.results}, we present results for the proton fraction in beta equilibrium and for the in medium neutrino cross sections.  Then, in Sec. \ref{sec.simulations}, we present preliminary simulations of the accretion phase in core collapse supernovae including $\Delta U$.  Finally, in Sec. \ref{sec.conclusions}, we discuss future work and conclude.

\section{Formalism}
\label{sec.formalism}
We consider a low density warm gas of only neutrons and protons.  For simplicity we neglect contributions of bound states involving three or more nucleons such as $^4$He nuclei.  They will be discussed below.  Note that the effects of deuteron bound states will be implicitly included in the virial coefficients, see below.  In the virial expansion the pressure $P$ is expanded in powers of the neutron $z_n$,
\begin{equation}
z_n=e^{\mu_n/T} 
\end{equation}
and proton $z_p$, 
\begin{equation}
z_p=e^{\mu_p/T}
\end{equation}
fugacities \cite{Horowitz:2005nd}.  Here the neutron chemical potential is $\mu_n$, the proton chemical potential is $\mu_p$ and $T$ is the temperature.  The virial expansion is valid at low densities and or high temperatures where $z_n$, $z_p<1$.  This roughly corresponds to densities $n$  so that $n/T^{3/2}<2\times 10^{-4}$\ fm$^{-3}$ MeV$^{-3/2}$.  The virial expansion further assumes that a phase transition has not taken place from the original very high temperature gas phase.

The pressure, to second order in the fugacities, is \cite{Horowitz:2005nd},
\begin{equation}
P = \frac{2T}{\lambda^3}\bigl\{ z_n + z_p + (z_n^2+z_p^2) b_n + 2z_pz_nb_{pn}\bigr\},
\label{Eq.p}
\end{equation}
where the nucleon thermal wavelength $\lambda$ is
\begin{equation}
\lambda=\Bigl(\frac{2\pi}{M T}\Bigr)^{1/2}
\end{equation}
with $M$ the nucleon mass.  We discuss the second order virial coefficients $b_n$ and $b_{pn}$ below.  For simplicity we neglect the mass difference between the neutron and proton.   This mass difference will contribute an effect similar to (but in general smaller than) the energy shift that we consider.  The neutron $n_n$ and proton $n_p$ densities are \cite{Horowitz:2005nd},
\begin{equation}
n_n = \frac{2}{\lambda^3}\bigl\{ z_n  + 2z_n^2 b_n + 2z_pz_nb_{pn}\bigr\},
\label{Eq.nn}
\end{equation}
\begin{equation}
n_p = \frac{2}{\lambda^3}\bigl\{ z_p  + 2z_p^2 b_n + 2z_pz_nb_{pn}\bigr\}\, .
\label{Eq.np}
\end{equation}
In general it is a simple matter to numerically find values of $z_p$ and $z_n$ that reproduce desired $n_n$ and $n_p$ values in Eqs.~\eqref{Eq.nn},\eqref{Eq.np}.  The chemical potentials then follow from $\mu_i=T$ln$(z_i)$.  Note that $\mu_i$ do not include the nucleon rest mass.

The virial coefficient $b_n$ describes pure neutron matter.  It is calculated from the observed isospin one nucleon-nucleon elastic scattering phase shifts.  We fit the numerical values from ref. \cite{Horowitz:2005nd} over the temperature range $1<T<20$\ MeV with,
\begin{equation}
b_n(T) \approx 0.3084 -\frac{0.0191}{T} + 5.8\times 10^{-4} T\, .
\label{Eq.bn}
\end{equation}
Here $T$ is in MeV.  Likewise the virial coefficient $b_{pn}$ describes the interactions between protons and neutrons.  We again fit the numerical values from ref. \cite{Horowitz:2005nd},
\begin{equation}
b_{pn}(T) \approx -0.9885 + 2.502\,\exp[\frac{2.099}{T}] -0.0179 T\, .
\label{Eq.bpn}
\end{equation}
Here the exponential term describes the contributions of deuterium bound states.  Equations \eqref{Eq.bn},\eqref{Eq.bpn} are good to 1\% over the range $1\ {\rm MeV} < T < 20\ {\rm MeV}$.  For higher temperatures  $20<T<50$\ MeV, Eqs.~\eqref{Eq.bn},\eqref{Eq.bpn}, although not explicitly fit to detailed phase shifts, still appear to provide reasonable qualitative behavior.    

\subsection{Energy Shift}
\label{subsec.energyshift}
To calculate the energy shift we start with the free energy density $f=(E-TS)/V$ with $E$ the internal energy, $S$ the entropy and $V$ the volume.  The use of the free energy, instead of the internal energy, will be discussed below.  In the virial expansion \cite{Horowitz:2005nd}
\begin{equation}
f=n_n\mu_n + n_p\mu_p - P\, .
\end{equation} 
Using Eq. \eqref{Eq.p} we have,
\begin{align*}
f=& {n_n T {\rm ln}z_n + n_p T {\rm ln} z_p} \\
& {- \frac{2T}{\lambda^3}\bigl\{z_n+z_p + (z_n^2+z_p^2)b_n+2z_pz_nb_{pn}\bigr\}}\, .
\end{align*}
Single particle energies will be calculated from $E_i=\partial f/\partial n_i$.  To simplify the calculation we invert Eqs.~\eqref{Eq.nn} and \eqref{Eq.np} to second order in the densities,
\begin{equation}
z_n\approx \frac{\lambda^3 n_n}{2} \bigl\{1 -\lambda^3(n_n b_n + n_p b_{pn})\bigr\}\, ,
\label{Eq.zn}
\end{equation}
\begin{equation}
z_p\approx \frac{\lambda^3 n_p}{2} \bigl\{1 -\lambda^3(n_p b_n + n_n b_{pn})\bigr\}\, ,
\label{Eq.zp}
\end{equation}
giving for the free energy density
\begin{multline}
f\approx n_n T {\rm ln}\frac{n_n\lambda^3}{2}+n_pT{\rm ln}\frac{n_p\lambda^3}{2}-T(n_p+n_n)\\
 \ \ \ -\frac{\lambda^3T}{2}\bigl\{(n_n^2+n_p^2)b_n+2n_nn_p b_{pn}\bigr\}+ O(n_i^3)\, .
\label{Eq.fn}
\end{multline}
This approximation is accurate at very low densities and will give us very simple results that provide physical insight.  Later we will obtain more accurate results by exactly solving Eqs.~\eqref{Eq.nn}, \eqref{Eq.np}.  It is now a simple matter to calculate the single particle energies using Eq.~\eqref{Eq.fn},
\begin{equation}
E_n=\Bigl(\frac{\partial f}{\partial n_n}\Bigr)_{n_p} = T{\rm ln}\frac{n_n\lambda^3}{2} - \lambda^3 T (n_n b_n+n_p b_{pn})\, ,
\label{Eq.en}
\end{equation}
\begin{equation}
E_p=\Bigl(\frac{\partial f}{\partial n_p}\Bigr)_{n_n} = T{\rm ln}\frac{n_p\lambda^3}{2} - \lambda^3 T (n_p b_n+n_n b_{pn})\, .
\label{Eq.ep}
\end{equation}
We measure the energy shift $U_i$ with respect to the energy of a noninteracting Fermi gas.  For a free Fermi gas one has second virial coefficients \cite{Horowitz:2005nd}, 
\begin{equation}
b_n^0=-\frac{1}{2^{5/2}}\, , \ \ \ \ \ \ \ \ \ \ b_{pn}^0=0\, .
\end{equation}
If one expands the pressure of a free Fermi gas in powers of the fugacity $z$, the coefficient of the $z^2$ term is given by $b_n^0$.  Therefore, the single particle energies $E_i^0$ for free Fermi gases are given by Eqs.~\eqref{Eq.en},\eqref{Eq.ep} with $b_n\rightarrow b_n^0$ and $b_{pn}\rightarrow 0$.
We now have simple results for the neutron and proton energy shifts,
\begin{equation}
U_n=E_n-E_n^0=-\lambda^3 T (n_n \hat b_n + n_p b_{pn})\, ,
\label{Eq.un}
\end{equation}
\begin{equation}
U_p=E_p-E_p^0=-\lambda^3 T (n_p \hat b_n + n_n b_{pn})\, ,
\label{Eq.up}
\end{equation}
and finally the difference in energy shifts is
\begin{equation}
\Delta U = U_n - U_p = \lambda^3 T(n_n-n_p)(b_{pn}-\hat b_n)\, .
\label{Eq.du}
\end{equation}
Here the difference in virial coefficients for interacting and free Fermi gases is
\begin{equation}
\hat b_n = b_n - b_n^0 = b_n +\frac{1}{2^{5/2}}\, .
\label{Eq.hatb}
\end{equation}
In the next Section we will show that the energy of the outgoing electron for $\nu_e$ capture on a neutron will be increased by $\Delta U$ while the energy of the outgoing positron from $\bar\nu_e$ capture on a proton will be decreased by $\Delta U$.  Equation \eqref{Eq.du} is a major result of this paper because it provides a model independent prediction for $\Delta U$ in terms of virial coefficients calculated from nucleon-nucleon elastic scattering phase shifts.

\subsection{Absorption cross sections}
\label{subsec.crosssection}
We now calculate the in medium cross section per unit volume for $\nu_e+n\rightarrow p + e$ \cite{Roberts:2012um}.
\begin{multline}
\frac{1}{V}\frac{d^2\sigma}{d{\rm cos}\theta dE_e}=\frac{G_F^2{\rm cos}^2\theta_C}{4\pi^2}\bigl[1+3g_a^2+(1-g_a^2){\rm cos}\theta\bigr]\\
E_e^2[1-f(E_e)] S_\nu(q_0,q)
\label{Eq.sigma}
\end{multline}
Here $G_F$ is the Fermi constant, $\theta_c$ the Cabibbo angle, $\theta$ the scattering angle, and the nucleon axial charge $g_a=1.267$.  The energy transferred to the medium is $q_0=E_\nu-E_e$ for neutrino energy $E_\nu$ and electron energy $E_e$, and $q$ is the momentum transferred to the medium $q^2=E_\nu^2+E_e^2-2E_\nu E_e{\rm cos}\theta$.  Finally, 
\begin{equation}
f(E_e)=\frac{1}{{\rm Exp}[(E_e-\mu_e)/T]+1}
\end{equation}
is the Fermi Dirac distribution for the outgoing electron.

Perhaps the simplest model for the response function $S_{\nu_e}(q_0,q)$ is to assume the neutrino strikes a heavy free nucleon at rest.  In this case the response function $S_{\nu_e}(q_0,q) \propto \delta(q_0)$ so that $E_e=E_\nu$.  As we will justify below, the effect of $\Delta U$ is to shift the response so that $S_{\nu_e}(q_0,q) \propto \delta(q_0+\Delta U)$.  In this case the energy of the outgoing electron will be 
\begin{equation}
E_e=E_\nu+\Delta U\, .
\end{equation}  
The ratio of the total cross section, angle and energy integral of Eq. \eqref{Eq.sigma}, $\sigma_{\nu_e}(\Delta U)$ calculated with $\Delta U$, to the cross section calculated with $\Delta U=0$, $\sigma_{\nu_e}(0)$ will be just the ratio of outgoing electron phase spaces,
\begin{equation}
\frac{\sigma_{\nu_e}(\Delta U)}{\sigma_{\nu_e}(0)}=\frac{(E_\nu+\Delta U)^2 [1-f(E_\nu+\Delta U)]}{E_\nu^2 [1-f(E_\nu)]}\, .
\label{Eq.rationu}
\end{equation}
This equation has a simple interpretation.  The shift $\Delta U$ increases the electron energy and this increases the available phase space and therefore the absorption cross section.

Likewise for antineutrino capture $\bar\nu_e+p\rightarrow n+e^+$ much the same thing happens in reverse.  Now the response would be approximately $S_{\bar\nu_e}(q_0,q)\propto \delta(q_0-\Delta U)$ so that the positron energy is reduced by the energy shift
\begin{equation}
E_{e^+}=E_{\bar\nu} - \Delta U\, .
\end{equation}
Therefore the ratio of total cross section $\sigma_{\bar\nu_e}(\Delta U)$ with $\Delta U$ to the cross section $\sigma_{\bar\nu_e}(0)$ without $\Delta U$ is
\begin{equation}
\frac{\sigma_{\bar\nu_e}(\Delta U)}{\sigma_{\bar\nu_e}(0)}=\frac{(E_{\bar\nu}-\Delta U)^2}{E_{\bar\nu}^2}\Theta(E_{\bar\nu}-\Delta U)\, .
\label{Eq.ratiobar}
\end{equation}
Now there is an energy threshold (given by the $\Theta$ function) where the cross section is approximately zero until $E_{\bar\nu} > \Delta U$.  In Eq.~\eqref{Eq.ratiobar} we have neglected Pauli blocking for the outgoing positron.  Again Eq.~\eqref{Eq.ratiobar} has a simple interpretation.  The shift $\Delta U$ reduces both the energy of the outgoing positron and the available phase space and this reduces the cross section for antineutrino absorption.   

We now wish to justify the simple results in Eqs.~\eqref{Eq.rationu},\eqref{Eq.ratiobar}
with a more detailed mean field model of the response function $S_{\nu_e}(q_0,q)$.  We start with a simple model for the in medium single neutron $\epsilon_n(q)$ and single proton $\epsilon_p(q)$ spectra,
\begin{equation}
\epsilon_n(q)=\frac{q^2}{2M} + U_n,\ \ \ \ \ \ \ \epsilon_p(q)=\frac{q^2}{2M}+U_p\, .
\label{Eq.spectrum}
\end{equation}
Here $M$ is the nucleon mass.  Note an effective mass $M^*$ in Eq.~\eqref{Eq.spectrum} is not expected to significantly change our results.  Furthermore at the low densities that we are interested in ($<0.001$ fm$^{-3}$), we expect $M^*\approx M$.  There is an important consistency requirement between the energy shift $U_i$ and the interacting chemical potential $\mu_i$.  The spectrum in Eq.~\eqref{Eq.spectrum}, with the interacting chemical potential, should give the correct nucleon density
\begin{equation}
n_i=2\int \frac{d^3p}{(2\pi)^3}\frac{1}{{\rm Exp}((\epsilon_i(p)-\mu_i)/T)+1}\, ,
\label{Eq.densityi}
\end{equation}
for $i=p$ or $n$.  This requires
\begin{equation}
U_i=\mu_i-\mu_i^f\, ,
\label{Eq.Uiconsistency}
\end{equation}
where $\mu_i^f$ is the chemical potential of a free Fermi gas.  Thus the energy shift is just the difference between the interacting and free chemical potentials.  If Eq. \eqref{Eq.Uiconsistency} is not satisfied, then the mean field response will likely be calculated for the wrong nucleon density.   Expanding Eq. \eqref{Eq.zn} for $\mu_n$ to second order in the density, and then calculating $\mu_n^f$ by replacing $b_n\rightarrow b_n^0$ and $b_{pn}\rightarrow 0$ one has
\begin{equation}
\mu_n-\mu_n^f=-\lambda^3T(n_n\hat b_n + n_p b_{pn}) + O(n_i^2)\, .
\label{Eq.mudif}
\end{equation}
This agrees with Eq. \eqref{Eq.un} to lowest order in the density.  Therefore calculating the energy shift in terms of the free energy, see Eq. \eqref{Eq.fn}, leads to a consistent definition of $U_i$.

We now calculate the neutrino response $S_{\nu_e}(q_0,q)$ for $\nu_e + n \rightarrow p + e$ in a mean field approximation assuming the spectra in Eq. \eqref{Eq.spectrum} \cite{Reddy:1998},
\begin{equation}
S_{\nu_e}(q_0,q)=\frac{M^2T}{\pi q(1-{\rm e}^{-z})}{\rm ln}\Bigl\{\frac{{\rm e}^{(e_{min}-\hat\mu_n)/T}+1}{{\rm e}^{(e_{min}-\hat\mu_n)/T}+{\rm e}^{-z}}\Bigr\}\, .
\label{Eq.response}
\end{equation}
Here, $e_{min}$ is
\begin{equation}
e_{min}=\frac{M}{2q^2}\bigl(q_0+\Delta U -\frac{q^2}{2M}\bigr)^2\, ,
\label{Eq.emin}
\end{equation}
and $\hat\mu_n=\mu_n-U_n$.  Finally, the detailed balance factor $z$ involves the energy transfer and the difference in chemical potentials $z=(q_0-\mu_n+\mu_p)/T$.  The response function for antineutrino absorption $S_{\bar\nu_e}(q_0,q)$ is obtained from Eq. \eqref{Eq.response} with $\Delta U\rightarrow -\Delta U$ and $\hat\mu_n\rightarrow \hat\mu_p$.  The cross section then follows from Eq. \eqref{Eq.sigma} with $E_e$ replaced by the positron energy $E_{e^+}$ and $f(E_e)$ replaced by the Fermi Dirac distribution for positrons.  In the next section we will show results for the total cross section obtained by integrating Eq. \eqref{Eq.sigma} using Eq. \eqref{Eq.response} over outgoing lepton energy and scattering angle.  These cross sections agree closely with the simple phase space ratios in Eqs. \eqref{Eq.rationu}, \eqref{Eq.ratiobar}.

\section{Results}
\label{sec.results}

In this section we present results for the composition of matter in beta equilibrium, neutrino and antineutrino absorption cross sections, and for the ratio of cross sections calculated with and without energy shifts.   We start with determining the proton fraction $Y_p$ for matter in beta equilibrium at baryon density $n$.  The procedure is to guess a $Y_p$ value and numerically solve Eqs.~\eqref{Eq.zn},\eqref{Eq.zp} with $n_p=Y_p n$ and $n_n=(1-Y_p)n$ for $\mu_p$ and $\mu_n$.  Next $Y_p$ is adjusted until 
\begin{equation}
\mu_n=\mu_p+\mu_e\, .
\label{Eq.betaeq} 
\end{equation}
Here $\mu_e$ is the chemical potential of a relativistic Fermi gas of electrons with density $n_p$.  Note that Eq.~\eqref{Eq.betaeq} assumes that the electron neutrino chemical potential is zero.  This is expected to be a good approximation near the neutrinosphere but need not be exactly true.

\begin{figure}[ht]
\begin{center}
\includegraphics[width=3.5in,angle=0] {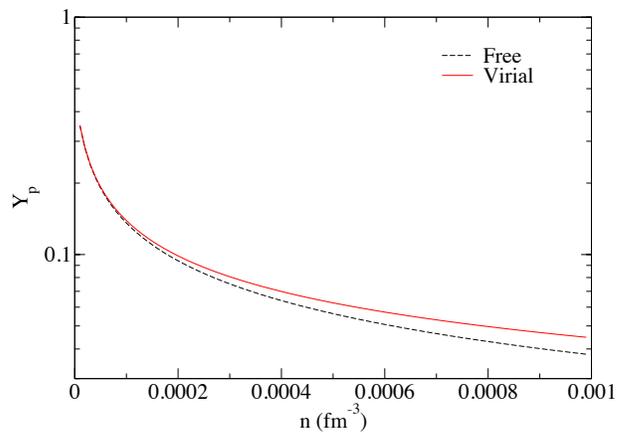}
\caption{(Color on line) Proton fraction $Y_p$ for matter in beta equilibrium at a temperature of $T=5$ MeV vs baryon density $n$.  The black dashed curve is for a free Fermi gas while the solid red line includes interactions in a virial expansion.}
\label{Fig1}
\end{center}
\end{figure}

In Fig.~\ref{Fig1} we show $Y_p$ in beta equilibrium versus baryon density $n$ for a temperature of $T=5$ MeV.   This temperature and range of densities in Fig. \ref{Fig1} are typical for the neutrinosphere, at least at early times.  We note that $n=0.001$ fm$^{-3}$ corresponds to a mass density of $1.7\times 10^{12}$ g/cm$^3$.  At later times, during protoneutron star cooling, the neutrinosphere may move to higher densities.  We discuss this more below.  We see that the virial interactions increase the equilibrium $Y_p$.

\begin{figure}[ht]
\includegraphics[width=3.7in,angle=0,clip=true] {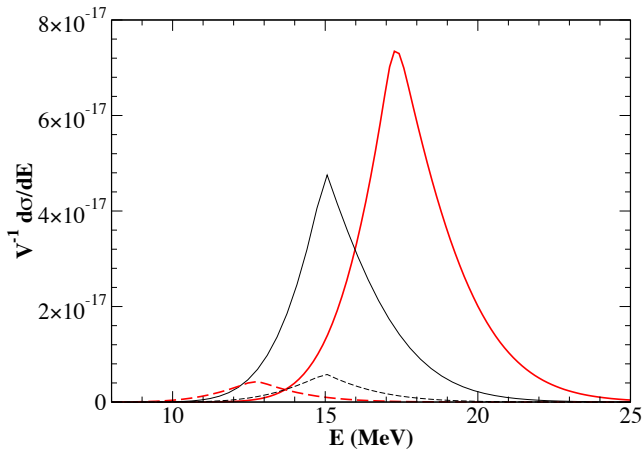}
\caption{(Color on line) Differential cross section per unit volume for neutrino (solid lines) or antineutrino (dashed lines) absorption versus energy $E$ of the outgoing charged lepton.  The thick red lines include energy shifts and peak at the lowest and highest energies.  Shifts are neglected for the thin black lines that peak near $E=15$ MeV.  This is for a temperature $T=5$ MeV, a density of $n=0.001$ fm$^{-3}$ and a neutrino (antineutrino) energy of 15 MeV.}
\label{Fig2}
\end{figure}

In Fig.~\ref{Fig2} we plot the differential cross section per unit volume $(1/V) d\sigma/dE$.  This is the angular integral of Eq. \eqref{Eq.sigma} at a density of $n=0.001$ fm$^{-3}$, $T=5$ MeV, and for a neutrino or antineutrino energy of $3T=15$ MeV.  This is a typical neutrino energy for this temperature.  Calculations that include energy shifts are plotted as thick red lines.  Note that the energy shifts $U_i$ are calculated from Eq. \eqref{Eq.Uiconsistency}.  This equation agrees with Eqs. \eqref{Eq.un}, \eqref{Eq.up} at low densities and is more consistent at higher densities as we discuss below.

Cross sections for antineutrinos in Fig.~\ref{Fig2} are smaller than neutrino cross sections simply because the density of protons $n_p$ is smaller than the density of neutrons $n_n$, see Fig.~\ref{Fig1}.  Note that the cross section is normalized per unit volume rather than per nucleon.  This reduction for antineutrinos is somewhat mitigated because positrons have less Pauli blocking than do electrons.

The energy shifts are seen in Fig.~\ref{Fig2} to increase the energy of the outgoing electron and to increase the cross section.  Likewise the energy shifts decrease the energy of the outgoing positron and reduce the antineutrino cross section.  We obtain the total cross section as the energy integral of $d\sigma/dE$.  The ratio of the total cross sections with and without energy shifts from Fig.~\ref{Fig2} agrees well with the simple phase space estimates of Eqs.~\eqref{Eq.rationu},\eqref{Eq.ratiobar}.

\begin{figure}[ht]
\includegraphics[width=3.7in,angle=0,clip=true] {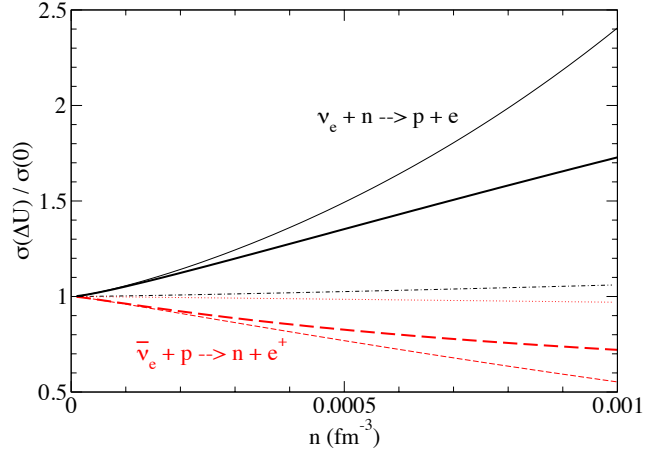}
\caption{(Color on line) Ratio of absorption cross sections with and without energy shift versus density $n$.  The solid black lines are for neutrinos, Eq. \eqref{Eq.rationu}, while the dashed red lines are for antineutrinos, Eq. \eqref{Eq.ratiobar}.  The thick lines use energy shifts from Eq. \eqref{Eq.Uiconsistency} while the thin lines use energy shifts valid to lowest order in the density from Eqs. \eqref{Eq.un}, \eqref{Eq.up}.    Finally, the dot-dashed (neutrino) and dotted (antineutrino) curves use the small mean field energy shifts from Eq. \eqref{Eq.umft}.  This is for $T=5$ MeV and for a neutrino energy of 15 MeV.}
\label{Fig3}
\end{figure}

The ratio of cross sections with and without energy shifts, Eqs.~\eqref{Eq.rationu},\eqref{Eq.ratiobar}, is plotted in Fig.~\ref{Fig3} for $T=5$ MeV and $E_\nu=15$ MeV.  The proton fraction $Y_p$ for all of the calculations is the beta equilibrium value in the virial expansion (solid red line in Fig.~\ref{Fig1}).  The thin lines use the lowest order energy shifts from Eqs.~\eqref{Eq.un},\eqref{Eq.up} while the thick lines use the difference in chemical potentials from Eq.~\eqref{Eq.Uiconsistency}.  These two prescriptions for the energy shifts agree at very low densities.  However, at higher densities the lowest order energy shifts predict a neutrino ratio that is larger than that for Eq.~\eqref{Eq.Uiconsistency}.  Since this later energy shift insures that the density is correct, we believe the thick lines in Fig.~\ref{Fig3} are the correct predictions for the neutrino and antineutrino ratios.

At a density of $n=0.001$ fm$^{-3}$ ($1.7\times 10^{12}$ g/cm$^3$) the neutrino absorption cross section is enhanced by 73\% because of the energy shifts, while the antineutrino absorption cross section is reduced by 28\%.  We now compare these virial expansion results to a mean field model.  For example, Roberts and Reddy \cite{Roberts:2012um} consider a mean field model GM3 that is normalized to nuclear phenomenology at saturation density $n_0=0.16$ fm$^{-3}$.  In this model the energy shift $\Delta U_{mf}$ is
\begin{equation}
\Delta U_{mf}=40\frac{n_n-n_p}{n_0}\ {\rm MeV}.
\label{Eq.umft}
\end{equation}
This model, compared to the virial expansion, predicts much smaller energy shifts.  Furthermore, the ratios of neutrino or antineutrino cross sections in Fig.~\ref{Fig3} are much closer to one .  For example at $n=0.001$ fm$^{-3}$, neutrino absorption is only enhanced by 6\%.  Alternatively, the mean field model IUFSU \cite{IUFSU} has a nonlinear density dependence and predicts an energy shift larger than that for GM3 but still less than the virial predictions.  In Table~\ref{table.one} we collect energy shifts for $n=0.001$ fm$^{-3}$ and $T=5$ MeV.  

\begin{center}
\begin{table}[h]
\begin{tabular}{|c|c|}
\hline
Model & $\Delta U$ (MeV)\\
 \hline
 Lowest order virial, Eq. \eqref{Eq.du}\ \ & 3.85 \\
 Virial $\mu_i-\mu_i^f$, Eq. \eqref{Eq.Uiconsistency}\ \ & 2.27\\
 Mean field model GM3, Eq. \eqref{Eq.umft}\ \ & 0.23\\
 Mean field model IUFSU \cite{IUFSU}\ \ & 1.11\\
 \hline
\end{tabular}
\caption{Energy shift $\Delta U$ predicted by different approaches at a density $n=0.001$ fm$^{-3}$ and a temperature $T=5$ MeV.}
\label{table.one}
\end{table}
\end{center}

The virial expansion calculates the pressure as a power series in the fugacities that is valid at low densities and or high temperatures.  For the conditions in Fig. \ref{Fig3}, the neutron fugacity is $z_n<0.16$ while the proton fugacity is $z_p<0.0045$.  These small values suggest that higher order corrections to the virial expansion will be small.

The inclusion of alpha particles was shown in ref. \cite{Horowitz:2005nd} to significantly enhance the symmetry free energy at higher densities and or lower temperatures.  Therefore we expect the formation of alpha particles (and other nuclei) to further enhance the virial energy shifts and neutrino absorption cross section changes.  Although preliminary calculations find small effects for the conditions in Fig. \ref{Fig3}, alpha particles do make significant contributions at lower temperatures and or higher densities.

In this paper we have only calculated the effects of an energy shift $\Delta U$ on neutrino interactions.  In addition, strong interactions will introduce other correlations between nucleons and these will impact neutrino interactions.  Often these correlations are included in model dependent RPA calculations, see for example \cite{Burrows:1999,Reddy:1999,Horowitz:2003}.  However the effects of correlations have been calculated for the long wavelength neutrino response of pure neutron matter using the virial expansion \cite{Horowitz:2006pj}.   In future work we will calculate the effects of correlations in nuclear matter using the virial expansion for both charged current and neutral current interactions.

\section{Supernova Simulations}
\label{sec.simulations}

We perform exploratory simulations to gauge the influence of the
energy shift on the neutrino signal during the accretion phase of
core-collapse supernovae.  We make use of {\tt nuGR1D}
\cite{oconnor:10,oconnor:12}, a spherically-symmetric,
general-relativistic, Eulerian hydrodynamics code with a two-moment
neutrino radiation transport solver.  For these simulations we take
the standard 15\,$M_\odot$, solar metallicity, core-collapse supernova
progenitor profile from \cite{woosley:95}. We employ the Lattimer and
Swesty \cite{lseos:91} equation of state with an incompressibility
modulus of 220\,MeV. The neutrino interaction rates are generated
using {\tt NuLib}, an open-source neutrino interaction library
available at {\tt http://www.nulib.org}. In {\tt NuLib}, the
absorption cross sections for neutrino and antineutrino capture on
free neutron and protons are taken from ref. \cite{burrows:06},
\begin{eqnarray}
\nonumber
\sigma_{\nu_e n}^\mathrm{abs} &= &\frac{G_F^2}{\pi}(1+3g_A^2)E_{e^-}^2\left(1-\frac{m_e^2}{E_{e^-}^2}\right)^{1/2}\\
&&\hspace*{0.5cm}\times W_M[1-f(E_{e^-})]\,,
\end{eqnarray}
and
\begin{eqnarray}
\nonumber
\sigma_{\bar{\nu}_e p}^\mathrm{abs} &= &\frac{G_F^2}{\pi}(1+3g_A^2)E_{e^+}^2\left(1-\frac{m_e^2}{E_{e^+}^2}\right)^{1/2}\\
&&\hspace*{0.5cm}\times W_M[1-f(E_{e^+})]\,,
\end{eqnarray}
where $W_M$ is the weak magnetism correction \cite{Horowitz:2002}, and
$f(E_{e^-})$ and $f(E_{e^+})$ are the Fermi Dirac distributions for
the outgoing electron and positron, respectively. We set $E_{e^-} =
E_\nu + \Delta_{np} +\Delta U$ and $E_{e^+} = E_\nu - \Delta_{np}
- \Delta U$.  Here, $\Delta_{np}$ is the neutron--proton mass
difference and $\Delta U$ is the energy shift. For simplicity in the
implementation, and to remain conservative in the calculation of the
energy shift, we use the following definition for $\Delta U$, which
follows closely from Eq.~\eqref{Eq.du},
\begin{equation}
\Delta U = T\,\,\mathrm{Min}(\lambda(T)^3(n_n-n_p),1)\,\,[b_{pn}(T^*) - \hat b_n(T^*)]\,,
\end{equation}
where the Min function is to limit the energy shift in regions where
the virial approximation breaks down. Likewise, we use $T^*
= \mathrm{Max}(T,1\,\mathrm{MeV})$ to limit the size of the virial
coefficients for low temperatures. The corresponding neutrino
emissivity from electron/positron capture on free
nucleons is calculated consistently using detailed
balance \cite{burrows:06,oconnor:12}.

\begin{figure}[ht]
\begin{center}
\includegraphics[width=0.98\columnwidth,angle=0] {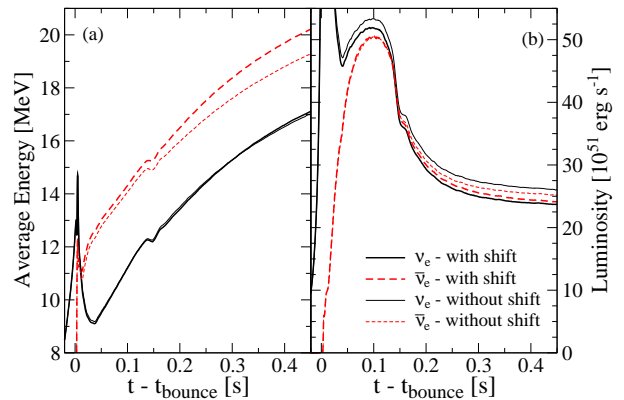}
\caption{(Color on-line) Average energy (a) and luminosity (b) of electron-type neutrinos and antineutrinos with and without energy shift. The solid black lines are for neutrinos, while the dashed red lines are for antineutrinos. The thick lines are from the simulation including the energy shift, while the thin lines do not include the energy shift. }
\label{sims}
\end{center}
\end{figure}

The overall shape of neutrino spectra look similar with and without the energy shift.  However the average energy and or luminosity can change.  In Fig.~\ref{sims}, we show the influence of the energy shift on the preexplosion neutrino signal through the evolution of the average
neutrino energy (a) and neutrino luminosity (b) for both the
electron neutrinos (solid, black lines) and antineutrinos (dashed, red
lines). The thick (thin) lines are the results with (without) the
energy shift. The third species included in the simulations, a
characteristic heavy-lepton neutrino, shows very little change with
the inclusion of the energy shift and is not shown. The energy shift
leads to an \emph{increase} in the average energy of the electron
antineutrinos. This is a result of a decrease in the absorption
opacity of the electron antineutrons on free protons, moving the
neutrinosphere for a given neutrino energy to lower radii and
therefore higher matter temperatures. The magnitude of this difference
is $\sim$0.25\,MeV, ($\sim1.8\%$) at 100\,ms after bounce. The
difference grows as the neutrinosphere recedes to higher densities
where the energy shift is large. After 450\,ms of postbounce
evolution, the difference is $\sim$0.94\,MeV ($\sim5\%$).  Note that neutrino-electron scattering can lower the $\bar\nu_e$ average energy and could somewhat limit the impact of $\Delta U$ on the $\bar\nu_e$ spectrum.  The average energy of electron neutrinos shows no dependence on the energy shift. The high free neutron fraction in the post shock region forces
the electron neutrino neutrinospheres to lower densities where the
influence of the energy shift is not as effective.  The other strong
effect of the energy shift is the \emph{reduction} of the electron
neutrino luminosity. The decrease is 1.5\,B\,s$^{-1}$ ($\sim3\%$) at
100\,ms after bounce and increases to 2.5\,B\,s$^{-1}$ ($\sim10\%$) at
450\,ms after bounce. We attribute the reduction to the lower electron
fraction found throughout the post shock region, but most importantly
around the neutrinospheres (where $\Delta Y_e/Y_e$ between the two
simulations is $\sim20\%$).  The lower electron fraction is a result
of the overall lower electron antineutrino opacities and emissivities
when the energy shift is included. We note that most ($\sim$90\%) of
the electron-neutrino luminosity change already obtains when including
the shift only in the antineutrino opacities.  Finally the small glitches visible in Fig. \ref{sims} at post bounce times near 0.15 seconds are from the advection of the silicon-oxygen interface through the shock (and are not caused by the energy shift).  These glitches are visible in previous simulations \cite{oconnor:12}.

The potentially significant quantitative changes that arise as a
result of the energy shift, both during the preexplosion phase (as
shown here) and presumably during the postexplosion phase warrant a
much more in-depth analysis in future work.  

\section{Discussion and conclusions}
\label{sec.conclusions}
In this section we discuss our results and conclude.  First, as shown in Fig.~\ref{Fig3}, the virial expansion energy shift $\Delta U$, at low densities, is much larger than that predicted by many mean field models and leads to much larger changes in neutrino cross sections. This is a major conclusion of the present paper.  Furthermore the virial predictions are based on elastic scattering phase shift data and are model independent.  Our preliminary simulations of the accretion phase of core collapse supernovae find that $\Delta U$ increases the $\bar\nu_e$ energies and decreases the $\nu_e$ luminosity, see Fig.~\ref{sims}.

We expect the energy shifts to have even larger effects on neutrino interactions at higher densities $n>0.001$ fm$^{-3}$.  However the virial expansion itself may not be directly applicable at these densities.   Therefore, it is important to calculate the properties of warm neutron rich matter in other microscopic approaches.  Although conventional quantum Monte Carlo approaches, such as ref. \cite{Gezerlis:2010,Gandolfi:2012}, may have difficulties with both the nonuniform matter and the high temperatures, other techniques may be more promising.  For example lattice effective field theory \cite{Lee:2009,Lee:2004} should be directly applicable.  Furthermore, Eq.~\eqref{Eq.Uiconsistency} shows that the energy shifts can be directly determined from chemical potentials.  Therefore one only needs to calculate the neutron and proton chemical potentials in a microscopic approach in order to determine the energy shifts and their impact on neutrino absorption.  

We emphasize that matter at neutrinosphere temperatures and sub-saturation densities can be directly produced in the laboratory with heavy ion collisions, see for example ref. \cite{Kowalski:2007}.  Furthermore, new radioactive beam facilities will allow the study of more neutron rich conditions. Terrestrial experiments can probe the equation of state, symmetry energy, and composition of neutrino sphere like matter.  These properties are important for neutrino interactions in core-collapse supernovae.  

Large energy shift effects could lower the energy or luminosity of $\nu_e$ radiated in supernovae.  Although about 20 $\bar\nu_e$ were detected from SN1987a, we have almost no experimental information on $\nu_e$ energies from SN1987a.  Therefore it is important to have a supernova $\nu_e$ detector with good energy resolution (such as liquid Ar \cite{ICARUS,LBNE}) to complement Super Kamiokande and other existing good $\bar\nu_e$ detectors.  We also note that Pb based detectors such as HALO \cite{HALO} have $\nu_e$ sensitivity, while the energy of $\nu_\mu$ and $\nu_\tau$ can be measured with neutrino-nucleus elastic scattering detectors \cite{nuelastic} for example.

A reduction in the $\nu_e$ energies or luminosity will likely make the neutrino driven wind above a protoneutron star more neutron rich.  This is important for nucleosynthesis. Perhaps energy shift effects are large enough to provide the necessary free neutrons in order for the r-process to occur in the neutrino driven wind.  Our preliminary simulations of the accretion phase of core collapse supernovae should be extended to the explosion and protoneutron star cooling phases using energy shifts that are calculated accurately at higher densities where the virial expansion is not directly valid.  We will present additional simulation results in a later publication. 

In conclusion, a proton in neutron rich matter is more tightly bound than is a neutron.  This energy shift $\Delta U$ increases the electron energy in $\nu_e+n\rightarrow p + e$, increasing the available phase space and absorption cross section, see Eq.~\eqref{Eq.rationu}.  Likewise $\Delta U$ decreases the positron energy in $\bar\nu_e+p\rightarrow n + e^+$, decreasing the phase space and cross section, see Eq.~\eqref{Eq.ratiobar}.  We have calculated $\Delta U$ using a model independent virial expansion and we find $\Delta U$ is much larger, at low densities, than the predictions of many mean field models.  Therefore $\Delta U$ could have a significant impact on charged current neutrino interactions in supernovae.  

We thank the INT for their hospitality during the program INT 12-2A when this work was started.  We thank Achim Schwenk for useful comments.  CJH is partially supported  by DOE grant DE-FG02-87ER40365.  The work of GS was supported by the DOE Topical Collaboration to study neutrinos and nucleosynthesis in hot dense matter.  CDO and EO are partially supported by NSF grants no. AST-0855535, OCI-0905046, and PHY-1151197.  Some of the numerical simulations were performed at Caltech's Center for Advanced Computing Research on the cluster ``Zwicky'' funded through NSF grant no.\ PHY-0960291 and the Sherman Fairchild Foundation.

\end{document}